\documentclass[reprint, showpacs, aps, nofootinbib, superscriptaddress, prb]{revtex4-1}
\usepackage{amsmath}
\usepackage{graphicx}
\usepackage{epstopdf}
\usepackage{hyperref}
\usepackage[utf8]{inputenc}
\usepackage[english]{babel}

\usepackage{soul}
\usepackage[usenames,dvipsnames]{color}
\usepackage[normalem]{ulem}

\begin{document}

\title{Possible quadrupolar nematic phase in the frustrated spin chain LiCuSbO$_4$: an NMR investigation}

\author{M. Bosio\v{c}i\'{c}}\email{marko.bosiocic@gmail.com}
\affiliation{Department of Physics, Faculty of Science, University of Zagreb, Bijeni\v{c}ka 32, HR-10000, Zagreb, Croatia}
\author{F. Bert}
\affiliation{Laboratoire de Physique des Solides, CNRS, Univ. Paris-Sud, Universit\'{e} Paris-Saclay, 91405 Orsay Cedex, France}
\author{S. E. Dutton}
\affiliation{Cavendish Laboratory, University of Cambridge, JJ Thomson Avenue, Cambridge CB3 0HE, United Kingdom}
\author{R. J. Cava}
\affiliation{Department of Chemistry, Princeton University, Princeton, New Jersey 08544, USA}
\author{P. J. Baker}
\affiliation{ISIS Facility, STFC Rutherford Appleton Laboratory, Didcot OX11 0QX, United Kingdom}
\author{M. Po\v{z}ek}
\affiliation{Department of Physics, Faculty of Science, University of Zagreb, Bijeni\v{c}ka 32, HR-10000, Zagreb, Croatia}
\author{P. Mendels}
\affiliation{Laboratoire de Physique des Solides, CNRS, Univ. Paris-Sud, Universit\'{e} Paris-Saclay, 91405 Orsay Cedex, France}

\begin{abstract}
The frustrated one-dimensional (1D) quantum magnet LiCuSbO$_4$ is one rare realization of the $J_1-J_2$ spin chain model with an easily accessible saturation field, formerly estimated to 12~T. Exotic multipolar nematic phases were theoretically predicted in such compounds just below the saturation field, but without unambiguous experimental observation so far. In this paper we present extensive experimental research of the compound in the wide temperature (30mK$-$300K) and field (0$-$13.3T) range by muon spin rotation ($\mu$SR), $^7$Li nuclear magnetic resonance (NMR) and magnetic susceptibility (SQUID). $\mu$SR experiments in zero magnetic field demonstrate the absence of long range 3D ordering down to 30mK. Together with former heat capacity data  [S.E. Dutton \emph{et al}, Phys. Rev. Lett. 108, 187206 (2012)], magnetic susceptibility measurements suggest short range correlated vector chiral phase in the field range $0-4$T. In the intermediate field values (5$-$12T), the system enters in a 3D ordered spin density wave phase with 0.75$\mu_B$ per copper site at lowest temperatures (125mK), estimated by NMR. At still higher field, the magnetization is found to be saturated above 13T where the spin lattice $T_1^{-1}$ relaxation reveals a spin gap estimated to 3.2(2)K. We narrow down the possibility of observing a multipolar nematic phase to the range 12.5$-$13T.

\end{abstract}


\maketitle

\section{INTRODUCTION}

The simple Heisenberg Hamiltonian describing an ideal one-dimensional (1D) spin $S=1/2$ system with only nearest-neighbor (NN) antiferromagnetic interactions was solved by Bethe in 1931. This remains until today one of the rare analytical solutions of such a Heisenberg Hamiltonian. Real materials also require the introduction of additional interactions, e.g.
next nearest neighbour (NNN) interaction, Dzyaloshinskii-Moriya (DM) interaction, effects of anisotropy etc. Analytical solutions, especially in the quantum case, often do not
exist, but the development of numerical methods has unveiled theoretical phase diagrams predicting the existence of novel exotic phases whose properties are not yet
completely understood, making 1D quantum spin systems a rich research topic in contemporary condensed matter physics.

In this vein, a special attention was recently given to phases which do not break time-reversal symmetry, {\it i.e.} $\langle S \rangle =0$
but are governed by spin tensor parameters hence named higher-rank multipolar-driven phase transitions, \cite{,,} the quadrupolar being the simplest. Such a spin-nematic phase
breaks O(3) spin symmetry but does not possess any magnetic moment.\cite{Penc2011}
It was first proposed in systems with onsite spin $S\geq 1$,\cite{Blume1969} and later a similar order was predicted in spin $S=1/2$ systems based on bond
order parameters.\cite{Grishchuk1984}
This could occur in frustrated Heisenberg spin $S=1/2$ chains under an applied field $h$, with a specific scheme of ferromagnetic (FM) nearest neighbor ($J_1$)
and antiferromagnetic (AFM) next  nearest neighbors interactions ($J_2$):\cite{Sato2013,Kecke2007,Sudan2009}

\begin{equation}
\label{Hamiltonian}
H=J_1\sum_{i}\textbf{S}_i\textbf{S}_{i+1}+J_2\sum_{i}\textbf{S}_i\textbf{S}_{i+2}-h\sum_{i}S_i^z.
\end{equation}
For systems with $J_1/J_2$ ratio in the range $[-2.7,0]$ three successive field-induced phases are predicted: a vector chiral (VC) phase in the low field region, a spin density wave (SDW$_2$)
phase of $2$ bound magnons in intermediate fields and a quadrupolar nematic (QN) phase above $\approx70\%$ of the saturation magnetization.\cite{Sudan2009}

The direct experimental  observation of a QN phase is demanding, since most local techniques are not sensitive to the dominant four point correlation function $\langle S^+_0 S^+_1 S^-_l S^-_{l+1}\rangle$, where $0,1,l$ and $l+1$ denote the positions of spins in a chain. Sato $et.$ $al.$\cite{Sato2009} suggested that NMR could indirectly probe the existence of the QN
phase through the temperature dependence of the spin-lattice relaxation rates $1/T_1$.

An ideal compound should have negligible interchain interactions to avoid 3D ordering and small intrachain interaction parameters $J_1$ and $J_2$  in order to reach the saturation of magnetization with $H$-$T$ parameters available in laboratories.
Among $J_1-J_2$ Heisenberg compounds, LiCu$_2$O$_2$\cite{Gippius2004}, Li$_2$CuO$_2$\cite{Giri2001}, Cs$_2$Cu$_2$Mo$_3$O$_{12}$\cite{Hoshino2013}, Rb$_2$Cu$_2$Mo$_3$O$_{12}$\cite{Hase2004}, LiCuVO$_4$\cite{Enderle2005}, PbCuSO$_4$(OH)$_2$\cite{Willenberg2012} and LiCuSbO$_4$, the three latter have been suggested as possible candidates to stabilize a QN phase in appropriate field conditions.
A recent 'tour de force' NMR study on LiCuVO$_4$\cite{Orlova2017} in pulsed high magnetic field up to 56 T pinpointed to a microscopic signature of QN phase,
but no $T_1$ measurements could be performed. For linarite PbCuSO$_4$(OH)$_2$, the stabilization of spin multipolar phases is now questioned by a recent neutron scattering experiment \cite{Cemal2017}. Finally, the recently discovered LiCuSbO$_4$ compound,\cite{Dutton2012} with the estimated coupling constants $J_1=-75\,$K and $J_2=34\,$K and rather low saturation field $\mu_0H_S\approx12-13\,$T, seems to be a very promising candidate for the realization of a QN phase: (i) No 3D ordering in zero field was observed down to $T=100$~mK which indicates that interchain couplings are negligible.
(ii) Neutron measurements together with susceptibility data show that short range correlations emerge below 9K, at temperatures depending on the magnitude of the applied magnetic field. (iii) An anomaly in the specific heat is detected for $T\sim 0.7\, \rm K$ which is enhanced on the application of a field and disappears for $\mu_0 H \geq 13$~T. One possible origin of this peak was suggested to be a multipolar transition in the vicinity of the saturation field.

In this paper we present an $H-T$ study of LiCuSbO$_4$ combining susceptibility, $\mu$SR and NMR measurements in the sub-Kelvin range, up to 13.3~T. As compared to a recent NMR study published in the course of our work~\cite{Grafe2017_2}, ours is performed on an {\it oriented} powder sample and at {\it temperatures down to} 30~mK. Through the lineshapes, we can monitor the various ground states under an applied field and we accurately determine the saturation field from the variation of the local magnetization determined through the line shift. We sketch a phase diagram combining our systematic analysis of NMR
spectra, $T_1$ relaxation times and $\mu$SR data, together with bulk magnetization measurements and former data from ref.[\onlinecite{Dutton2012}]. Our analysis shows that LiCuSbO$_4$ is indeed a very good candidate for the realization of a QN phase in the field range between 12.5 and 13 T. Our results partly differ from ref.[\onlinecite{Grafe2017_2}]. and are discussed with respect to that work.

The paper is structured as follows: in section II, we give the details about the sample preparation, the structure and the procedure used to orient the NMR sample. In section III we present our $\mu$SR results in zero field down to 30mK. Section IV gives our SQUID experiment results from low field to 7T, down to 300mK. The NMR spectral analysis is presented in section V, where we also
discuss the saturation field value which is crucial for the determination of the field range where a QN phase can be expected. Field and temperature dependent NMR spin lattice relaxation $T_1^{-1}$ measurements are presented in section VI. In the discussion we compare the compound with other possible 1D candidates described by the $J_1-J_2$ Hamiltonian and finally build the $H-T$ phase diagram from all
available experimental data on LiCuSbO$_4$.

\section{Crystal structure and sample preparation.}
\begin{figure}
\includegraphics[width=88mm]{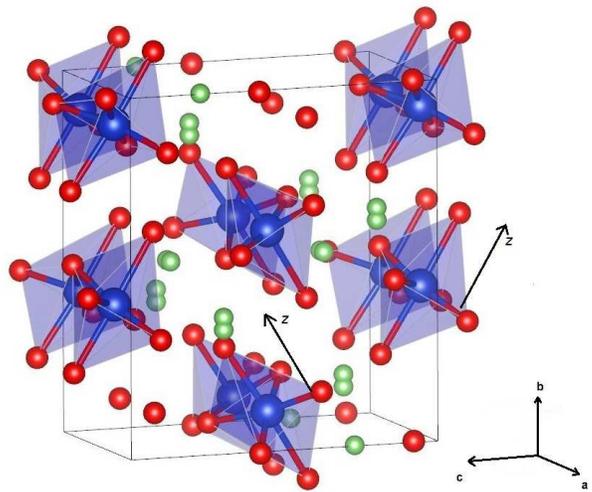}
\caption{(Color online) Crystal structure of LiCuSbO$_4$. Unpaired spins are localized on Cu$^{2+}$ ions (Blue) centered in CuO$_6$ octahedra. Edge sharing octahedra form spin chains parallel to the $a$ axis. The local $z$ axes of two adjacent spin chains are canted by $34^\circ$ in opposite directions with respect to the $b$ axis. Distribution of Li(1) (Green) is supposed to be in the direction of $b$ axis, while Li(2) in the direction of $a$ axis. Oxygen atoms are labeled in red color. Antimony ions are omitted for clarity.}
\label{structure}
\end{figure}
LiCuSbO$_4$  crystallizes in the space group $Cmc2_1$ with tetragonal unit cell ($a=5.74260\,$\AA, $b=10,86925\,$\AA, $c=9.73048\,$\AA). The copper ions Cu$^{2+}$ carry $s=1/2$ spins. The structure is shown in Fig.~\ref{structure}\cite{Momma2011}. The basic building blocks are edge-sharing CuO$_6$ octahedra which form chains along the $a$ axis. The Cu-O bonds perpendicular to the $a$ axis are significantly elongated, defining local $z$ axes for each chain as marked in Fig.~\ref{structure}.  The $z$ axes of the CuO$_6$ octahedra in adjacent chains are not parallel, but canted $34^{\circ}$ in opposite directions with respect to $b$ axis. The unit cell contains two crystallographic positions of lithium, Li(1) and Li(2), slightly distributed along $b$ axis and $a$ axis, respectively. Crystallographically, this is modeled as two close sites with 50\% occupancy.\cite{Dutton2012}
Polycrystalline samples were synthesized as explained in 
ref.~\onlinecite{Dutton2012}.  For the magnetization and NMR measurements, the polycrystalline powder was ground and mixed with an epoxy glue and during the hardening, the sample was rotated around an axis perpendicular to a magnetic field $\approx 7$T for 24h. This procedure ensures that the hard magnetization axis of the sample aligns with the rotation axis. The orientation of the powder sample was checked by NMR as described in Appendix A. Further X ray diffraction refinements on the oriented sample have shown that the rotation axis is perpendicular to the $a$ axis, i.e. the hard axis lays in the $bc$ plane.
In NMR and magnetization measurements the magnetic field was applied along the hard axis, perpendicular to the direction of the chains ($a$ axis).

\section{$\mu$SR Measurements}

In previous heat capacity measurements\cite{Dutton2012}, no magnetic transition was observed down to $100\,$mK. However signatures of spin-glass like states or subtle spin freezing involving only part of the spin degrees of freedom are often elusive in such measurements. In order to confirm the nature of the zero field ground state on a firmer ground, we performed $\mu$SR measurements  which are extremely sensitive to even tiny internal fields~\cite{Blundell1999}. The measurements were performed at the ISIS pulsed muon source facility (UK) on the MUSR spectrometer.  We used a powder sample $\approx 1\,$g from the same batch as for the NMR measurements. The loose powder sample was spread on a silver holder plate and covered with drops of diluted GE-varnish to ensure a good thermal contact. A helium-bath cryostat with a dilution refrigerator insert was used for temperature control in the range $0.03\,$K to $10\,$K.

\begin{figure}
\includegraphics[width=88mm]{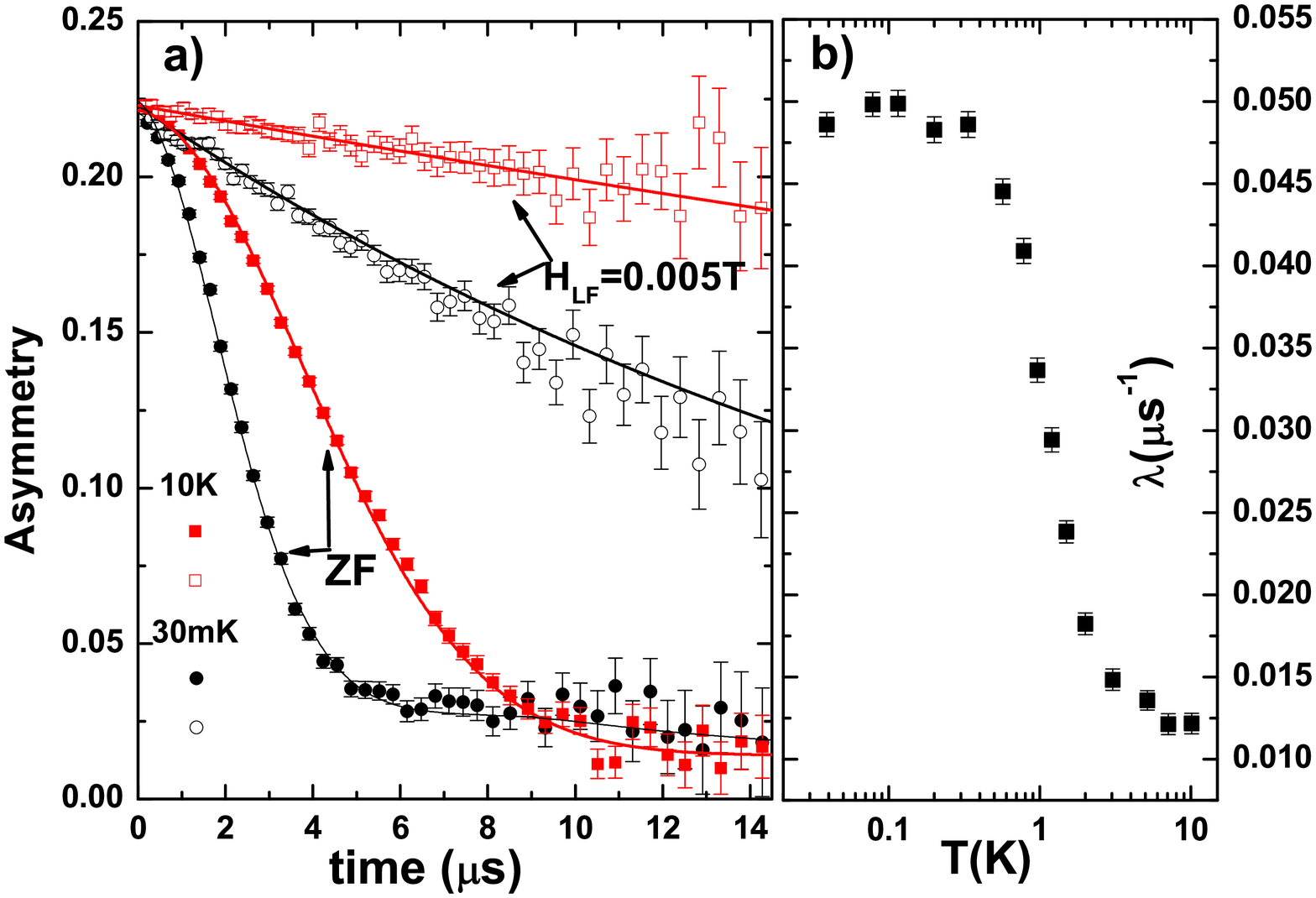}
\caption{ (Color online) a) Asymmetries of the muon decay in zero field (solid symbols) and with a small $5\,$mT longitunal field (open symbols) at $30\,$mK and $10\,$K. Lines are fits to the data (see main text). b) Temperature dependence of the muon relaxation rate extracted from exponential fits of the muon decay asymmetry measured with a $5\,$mT applied longitudinal field. }
\label{musrfig}
\end{figure}

Muon decay asymmetries recorded at the highest and base temperatures of the experiment are shown in Fig.~\ref{musrfig}. At high temperature the zero field relaxation is dominated by the rather strong Cu and Li nuclear moments which are disordered and static on the muon time scale. The initial gaussian shape is characteristic of the static Kubo Toyabe relaxation $G(t,\Delta,0)$ expected for frozen random fields with a distribution width $\Delta$~\cite{Yaouanac2011,Uemura1998}. However, the ''1/3rd'' recovery at long times of the Kubo Toyabe function is not observed here, likely because of numerous muon stopping sites and nuclear moments sizes.  In addition to nuclear magnetism, the fast fluctuating electronic moments contribute to a weak exponential relaxation which can be isolated by applying a small longitudinal field of $5\,$mT to decouple the nuclear static relaxation as shown in Fig.~\ref{musrfig}. The $10\,$K asymmetries were thus fitted to the model
\begin{equation}
a(t)=a_0P(t)\exp(-\lambda t)+a_b
\label{musrequation}
\end{equation}
with $P(t)=\exp(-\sigma^2 t^2)$ in zero field and $P(t)=1$ for a $5\,$mT applied field. The $T$-independent parameters $a_0=0.220(5)$ and $a_b=0.014(5)$ stand for the initial muon decay asymmetry and the fraction of muons stopping in the silver sample holder. From the gaussian decay $\sigma=0.18(1)\, \mu$s$^{-1}$ and the muon gyromagnetic ratio $\gamma_\mu = 2 \pi \times 135.54\,$MHz/T, we estimate the frozen field distribution width $\Delta \approx \sigma / \gamma_\mu =0.21(1)\,$mT consistent with nuclear magnetism.

Upon cooling the relaxation increases gradually down to $0.03\,$K where the zero field asymmetry shows a steeper gaussian decay and crosses the $10\,$K one around $9~\mu$s. The crossing indicates that some muons experience static fields along their initial polarization yielding a non relaxing asymmetry -the so-called ''1/3rd'' tail in a fully magnetic powder sample. This is therefore a clear evidence for the presence of frozen internal fields at base temperature, besides the nuclear ones which are temperature independent. However and most remarkably, these static internal fields are surprisingly small since they produce a relaxation hardly faster than the $10\,K$ one due to nuclear fields. To quantify these model-free observations, we fitted the $0.03\,$K data to Eq.~\ref{musrequation} (see black solid lines in Fig.~\ref{musrfig}). For the zero field asymmetry we used $P(t)=G(t,\Delta,\nu)$, the dynamical Kubo-Toyabe relaxation which accounts for essentially static internal fields with $\Delta=0.42(2)\,$mT, slowly fluctuating at the rate $\nu=0.26(1)\,$MHz $\ll \gamma_\mu \Delta$.
For comparison, one can estimate the level of spin ordering as follows. Making the usual assumption that muons stop close to negatively charged O$^{2-}$ ions, $\approx 2$\AA$\,$away from the Cu$^{2+}$ sites, in the hypothetical case of full ordering with $1~\mu_B$ per Cu ion, one would expect dipolar local field $\mu_0\mu_B/(4\pi r^3)\approx 0.12\,$T at the muon sites. The observed local fields are three orders of magnitude lower and one can discard standard magnetic phases with on-site frozen moments as the zero field ground state. It is possible that the marginal spin freezing reflects the existence of short range correlations as suggested in Ref. [\onlinecite{Dutton2012}].

Since the static internal fields remain weak in the whole $T$ range, the dynamical relaxation channel can be probed in isolation down to $0.03\,$K by applying the same small $5\,$mT longitudinal field as at $10\,$K. The relaxation remains exponential at all $T$. The temperature dependence of the $\lambda$  parameter of Eq.~\ref{musrequation} is shown in Fig.~\ref{musrfig}.b). The increase of $\lambda$ signals a moderate slowing down of the spin dynamics as if the system were approaching a transition around $1\,$K. However, instead of diverging and decreasing below $1\,$K, $\lambda$ levels off showing the persistence of slow fluctuations in the ground state. The absence of a peak in $\lambda(T)$ is another evidence for the absence of 3D ordering or even standard spin-glass freezing in LiCuSbO$_4$ down to 0.03~K.

\section{Low temperature magnetic susceptibility}

\begin{figure}
\includegraphics[width=88mm]{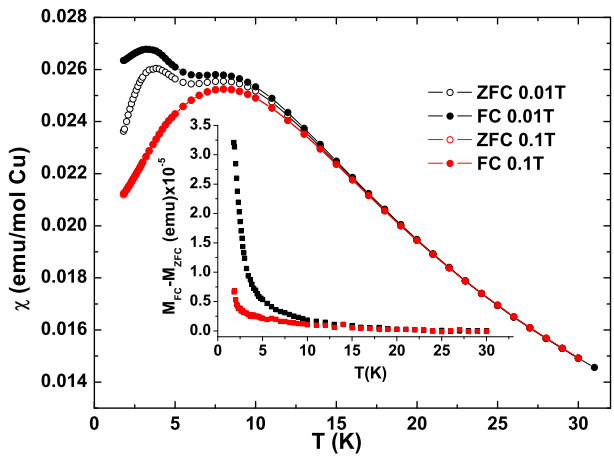}
\caption{ (Color online) Low field susceptibility measurement on a non oriented powder sample. In the magnetic field of 0.01T, FC and ZFC susceptibility start to differ at 15K. Upon increasing magnetic field this difference vanishes. Inset show differences between between FC and ZFC magnetization curves in a 0.01T (black) and 0.1T (red).}
\label{squid}
\end{figure}

The magnetization of a powder sample and an oriented powder sample, prepared as described in section II, was measured  versus temperature in various magnetic fields in a commercial MPMS-7 Quantum Design SQUID magnetometer equipped with a $^3$He insert for subkelvin measurements.
In the lowest investigated field of $0.01\,$T, a small difference between the field cooled and zero field cooled susceptibilities is observed below about 15K (see Fig.~\ref{squid}).
This irreversibility is very weak in $0.1\,$T and disappears upon applying higher fields. The occurrence of some level of spin freezing in low field is consistent with the $\mu$SR data.

\begin{figure}[b!]
\includegraphics[width=88mm]{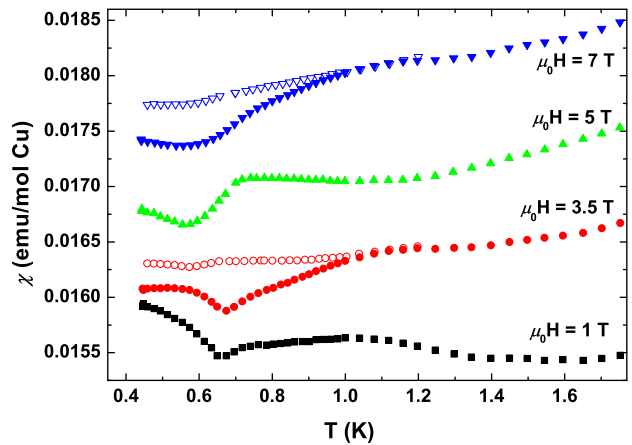}
\caption{(Color online) Low-temperature magnetic susceptibility of the oriented powder LiCuSbO$_4$ with magnetic field parallel to the average hard axis (full symbols). The curves are shifted vertically for clarity. Empty symbols: for comparison, susceptibility of the unoriented powder sample at 3.5 and 7T.}
\label{Chi_lowT}
\end{figure}

We have further measured the low-temperature magnetic susceptibility of an oriented powder sample in four characteristic fields (Fig.~\ref{Chi_lowT}) applied along the hard axis, perpendicular to the direction
of the chains. All the measured susceptibility curves show some feature at $T=0.7\, \rm K$, indicating the existence of a phase boundary. In the lower fields ($1\,\rm T$ and $3.5\,\rm T$), the transition is defined by a sharp minimum, while at higher fields ($5\,\rm T$ and $7\,\rm T$) an increase of $\chi$ occurs at the same temperature. These different features suggest that the low-temperature phases are distinct, i.e. there exist a phase boundary, when the field is inscreased at constant temperature, in between $3.5\,\rm T$ and $5\,\rm T$. Similar small changes in the signatures of the transitions were observed in single crystal of linarite PbCuSO$_4$(OH)$_2$ and used to draw its complex magnetic phase diagram.\cite{Willenberg2012}. For comparison, the transitions in LiCuSbO$_4$ are even weaker and hardly detectable in an unoriented powder sample (empty symbols in Fig.~\ref{Chi_lowT}).

\section{NMR spectra and determination of the saturation field}

\begin{figure}[b!]
\includegraphics[width=88mm]{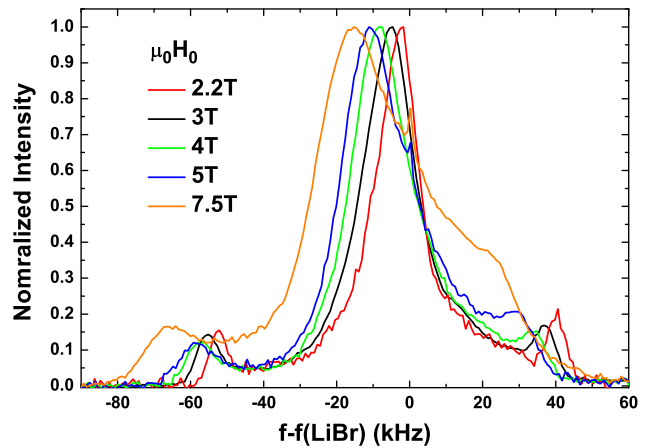}
\caption{(Color online) Room temperature $^7$Li spectra in various magnetic fields (up to $7.5\, \rm T$). At low fields one can clearly distinguish the quadrupolar satellites of the Li site with the highest quadrupolar coupling.} 
\label{fieldRT}
\end{figure}

NMR measurements were performed on $^7$Li nuclear spin $I=3/2$ ($\gamma / 2\pi=16.54607\,$MHz/T). The spectra were obtained using a solid echo sequence $\pi/2-\tau-\pi/2$, with a pulse length between $1.5\,\mu$s and $3\,\mu$s when the helium flow cryostat was used and $10\,\mu$s in the dilution fridge environment. A solution of LiBr in deionized water was used to determine the NMR reference frequency $\nu_0$.

\begin{figure*}
\includegraphics[width=160mm]{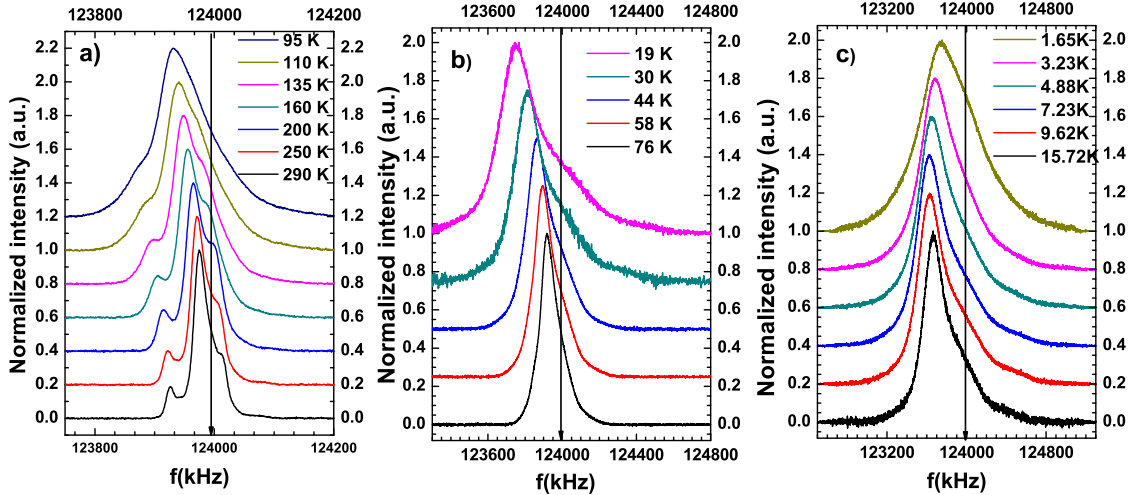}
\caption{(Color online) Spectra at $\mu_0H=7.5\,$T for different temperatures. The vertical lines denote the reference frequency $\nu_0$.}
\label{spectraT}
\end{figure*}

Spectra at room temperature in magnetic fields up to 7.5~T are shown in Fig.~\ref{fieldRT}.  At low fields, below $5\,$T, the central lines from the two crystallographic lithium sites Li(1) and Li(2) overlap and a pair of quadrupolar satellites with $\nu_Q=47(1)$kHz is clearly resolved. Since the pulse length was optimized for the central transition, the quadrupolar satellites are less excited, and the theoretical ratio 3:4:3 for the central line and satellites intensities is not obeyed.  As the field is increased, these central lines of the two Li sites start to differ due to different hyperfine coupling constants and the central line coming from the less coupled one starts to overlap with the right quadrupolar satellite of the other.

The evolution of the spectral shape at $7.5\,$T with decreasing temperature can be followed in Fig.~\ref{spectraT}. The spectral line shifts to lower frequencies and simultaneously broadens magnetically. Eventually, the quadrupolar satellites cannot be resolved anymore below about $80\,$K.  The broad NMR spectra are obtained by combining partial spectra measured at periodically spaced irradiation frequencies.

\begin{figure}
\includegraphics[width=88mm]{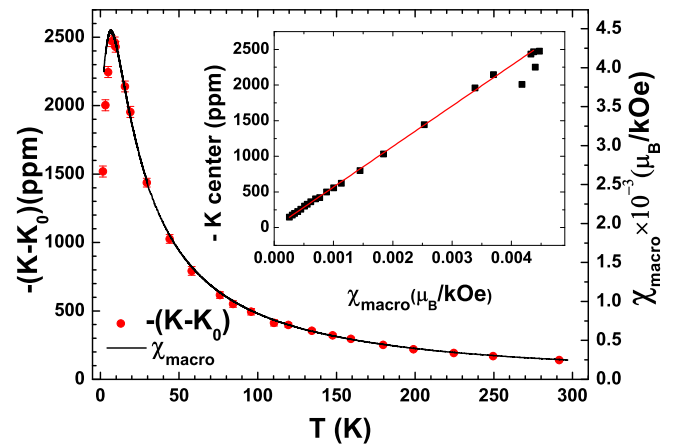}
\caption{(Color online) Temperature dependence of the NMR shift (red circles) and macroscopic SQUID susceptibility (line). Inset: The Jaccarino plot shows the dependence of the NMR shift on the macroscopic magnetic susceptibility measured at 8~T.}
\label{jaccarino}
\end{figure}

\begin{figure}
\includegraphics[width=88mm]{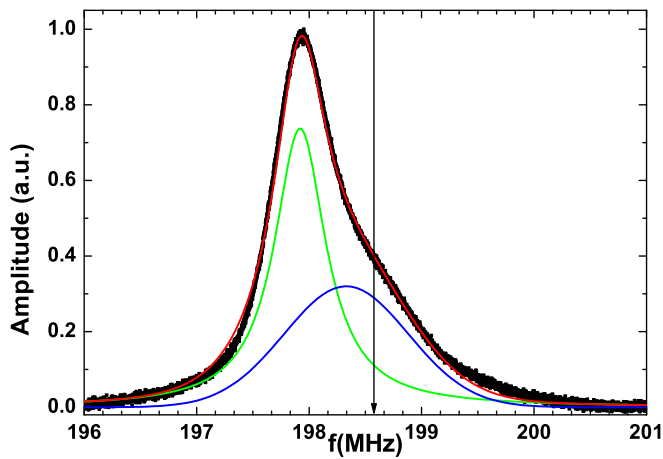}
\caption{(Color online) Spectrum recorded at $T=1.55\,$K in magnetic field $\mu_0H_0=12\,$T (black). The red line is a simulation composed of a Lorentzian (green) and Gaussian (blue) line. Ratio of integrated intensities is 55:45 in favor of the Lorentzian line. The contribution of the Gaussian line at the main peak position is $\approx28\%$. The vertical line shows the shift reference.}
\label{spectra12T}
\end{figure}

To estimate the strength of the coupling between the $^7$Li nuclei and the Cu electron spins,  we compare the spectral shift $K=(\nu-\nu_0)/\nu_0$ to the macroscopic magnetic susceptibility $\chi_{macro}$ measured at similar fields, respectively $7.5\,$T and $8\,$T (see Fig.~\ref{jaccarino}), using the general relation
\begin{equation}
K=A_{hf}\chi_{macro}+K_0 \ \ ,
\label{EqKvsChi}
\end{equation}
where $K_0$ is the temperature independent orbital shift, and $A_{hf}$ is the hyperfine coupling, assumed to be isotropic. Although there are two Li lines corresponding to the two Li sites, with different hyperfine coupling constants, we could only track reliably, on the whole temperature range, the shift of the maximum of the spectra corresponding to the most coupled Li site (see below Fig. \ref{spectra12T}).
As shown in the Jaccarino plot in the inset of Fig.~\ref{jaccarino}, Eq.~\ref{EqKvsChi} is satisfied on a large $T$ range down to about $9\,$K which yields the hyperfine coupling constant  $A_{hf}=-0.57(2)$ kOe/$\mu_B$ and $K_0=0 (6)$ppm.

Below $9\,$K, the NMR shift and the macroscopic susceptibility start to differ slightly. This difference may be due to the fact that the NMR shift was measured on an oriented powder sample, while the SQUID data were measured on an unoriented one. Furthermore, even a tiny amount of impurities with Curie-like contributions, can alter the macroscopic susceptibility at low temperature while the NMR shift at peak position remains unaffected. The smallness of the difference between the two measurements, both showing clearly a maximum, supports the high quality of the sample and the claim for an upper limit of the impurity concentration, such as Li/Cu or Sb/Cu intersite mixing, of 1\%~\cite{Dutton2012}.

When the magnetic moment increases (by applying a higher magnetic field or by lowering the temperature), the typical spectral shape resembles the one shown in Fig.~\ref{spectra12T} ($\mu_0H_0=12\,$T at $1.55\,$K ). The spectrum is well simulated by a Lorentzian line for the strongly coupled site and a Gaussian line for the weakly coupled one. The ratio of their intensities is 55:45 in favor of the Lorentzian line.  The contribution of the weakly coupled site to the amplitude of the main peak is $\approx28\%$.

\begin{figure}[h!]
\includegraphics[width=88mm]{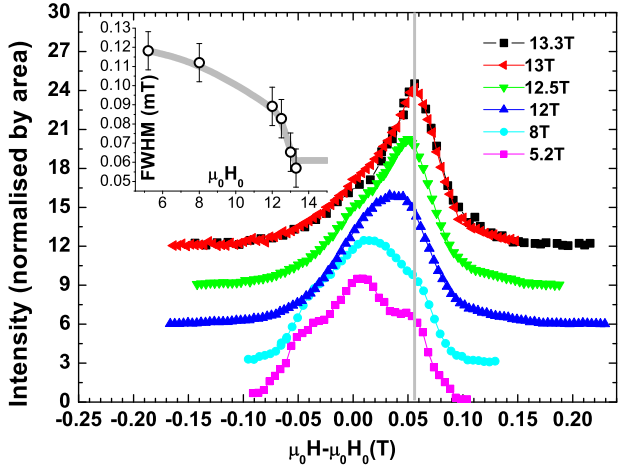}
\caption{ (Color online) Spectra obtained by field sweeps at different frequencies ($\nu=\gamma\mu_0H_0$) at $T=0.125\,$K. The vertical line represents the maximum hyperfine field of $0.057\,$T corresponding to 1$\mu_B$ per Cu site. In $5.2\,$T and $8\,$T spectra are squarish in shape indicating spin freezing. As the field increases, the system polarizes and the spectra narrow.  The inset shows the full width at half maximum (FWHM) of the spectra in the main panel. The grey line is a guide to the eyes.}
\label{DRspectra}
\end{figure}

At lower temperatures, below $1\,$K, the spectral shape considerably changes and becomes strongly field dependent as shown in Fig.~\ref{DRspectra}. The vertical line in that figure denotes the maximum shift of $0.057\,$T using the hyperfine constant and assuming a full 1$\mu_B$ moment per Cu site. The spectral widths at $0.125\,$K are substantially larger than the one at $1.55\,$K in the corresponding fields. These broad spectra are obtained at fixed irradiation frequencies $\nu_0$ by field sweeps around the reference field $\mu_0H_0=2\pi \nu / {\gamma}$.
The squarish  shape of the $5.2\,$T spectrum is characteristic of the powder average in a magnetically ordered system where all nuclei experience the same magnitude of the hyperfine field.  From the spectral width one can estimate the amplitude of the onsite Cu moment to be $0.75\mu_B$. The narrower component on top of the squarish one can likely be assigned to the less coupled site.
In stronger applied fields, the spectra become narrower (inset in Fig.~\ref{DRspectra}  ) and shifted towards higher fields. This is the expected consequence of the increasing polarization of the system towards saturation when all the spins point in the same direction along the applied field. The spectra at $13\,$T and $13.3\,$T overlap indicating that  the system is fully polarized above $13\,$T. The shift value for the spectra at $13\,$T and $13.3\,$T matches perfectly the maximum shift value calculated independently from the determination of the hyperfine constant at high temperatures (vertical line in Fig. \ref{spectra12T}).

\begin{figure}[t!]
\includegraphics[width=88mm]{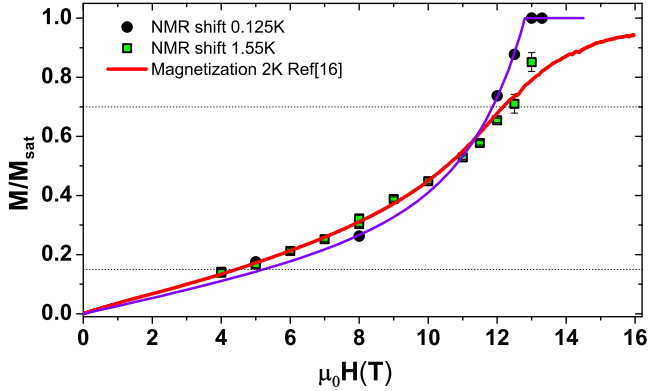}
\caption{ Field dependence of the NMR peak position at $T=0.125\,$K (black circles) and $T=1.55K$ (green squares) compared to the bulk magnetization at 2~K from ref.~\onlinecite{Dutton2012} (red line). The purple line is a guide to the eyes for the $T=0.125\,$K NMR data.}
\label{saturacija}
\end{figure}

The saturation field value of LiCuSbO$_4$ was formerly estimated from bulk magnetization data at 2~K on a powder sample (full red line in Fig.~\ref{saturacija})~\cite{Dutton2012}. The full saturation was not reached for the highest $16\,$T available field although an inflection occurs roughly between 12 and $13\, \rm T$. The magnetization curve could be tentatively reproduced by introducing exchange anisotropy but finite temperature effects, powder averaging and possible impurity contribution preclude a precise determination of the saturation field which is extremely important in the search for the QN phase.

An alternative way to estimate the saturation field is the measurement of the local magnetization at $^7$Li site through the position of the NMR line shift (Fig.~\ref{DRspectra}), with the advantage of using an oriented powder sample.
The field dependence of the NMR shift at $T=0.125\, \rm K$ is shown by full circles in Fig.~\ref{saturacija}. It is clear that the magnetization is still not saturated at $12.5\,$T, but that it is saturated at $13\,$T. Compared to the bulk magnetization data taken at $T=2\, \rm K$, the increase of the NMR shift is much steeper around $12\, \rm T$.

To check the effect of temperature, we have also measured NMR shifts at $T=1.55\, \rm K$ (squares). The data correspond better to the bulk magnetization at $2\, \rm K$, despite the sample orientation for the NMR measurements. In conclusion, we estimate the saturation field of LiCuSbO$_4$ to be larger than $12.5\,$T and slightly below $13\,$T.
Our results emphasize the advantage for the determination of the saturation field to investigate the local magnetization at very low temperatures, preferably on oriented powder.

%

%
%

\section{Relaxation measurements}

\begin{figure}[t!]
\includegraphics[width=88mm]{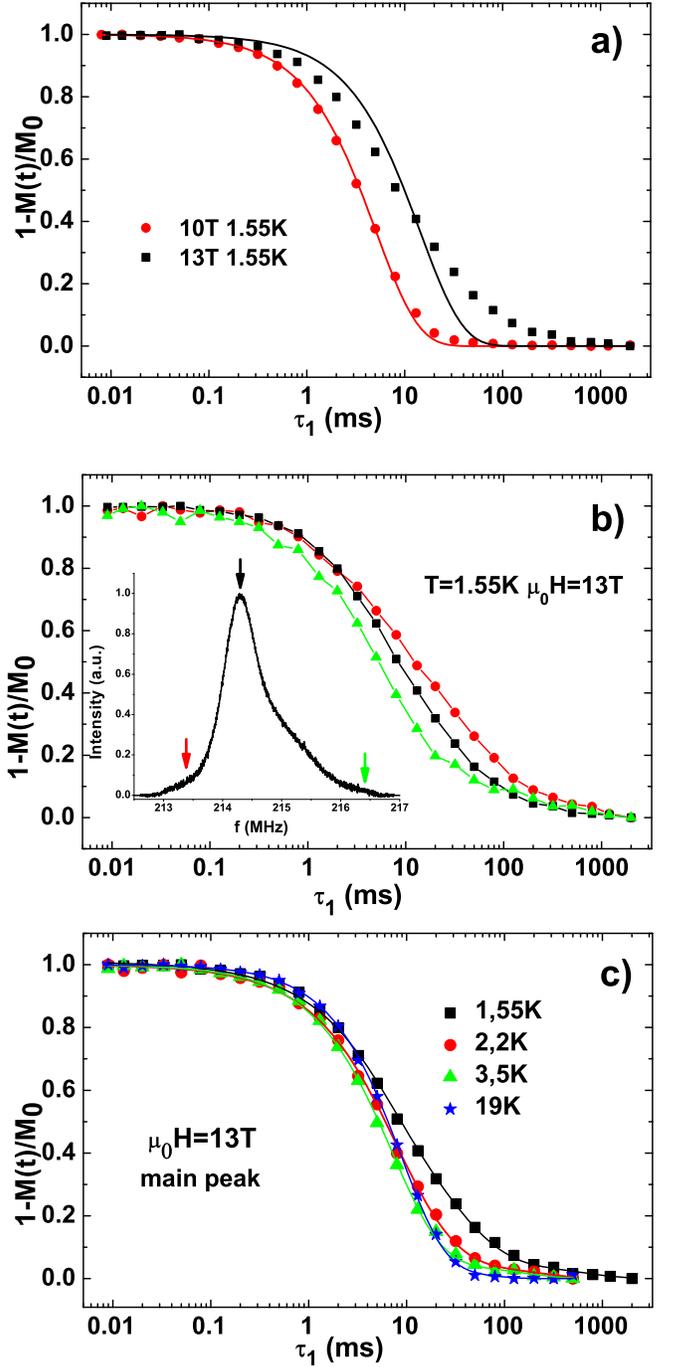}
\caption{ (Color online) a) Spin-lattice relaxation (central peak) measured at $T=1.55\,$K at $\mu_0 H=10\,$T (red circles) and at $\mu_0 H=13\,$T (black squares). The lines are single-exponential fits;
b) Relaxation of the magnetization at different parts of spectra at the temperature $T=1.55$K. Arrows on the spectra correspond to frequencies where relaxation was measured. Lines are  guide for the eyes;  c) Relaxation curves at the main peak at various temperatures. Full lines are obtained by the fit to Eq. (\ref{relaksacija}).}
\label{curves}
\end{figure}

\begin{figure}[t!]
\includegraphics[width=88mm]{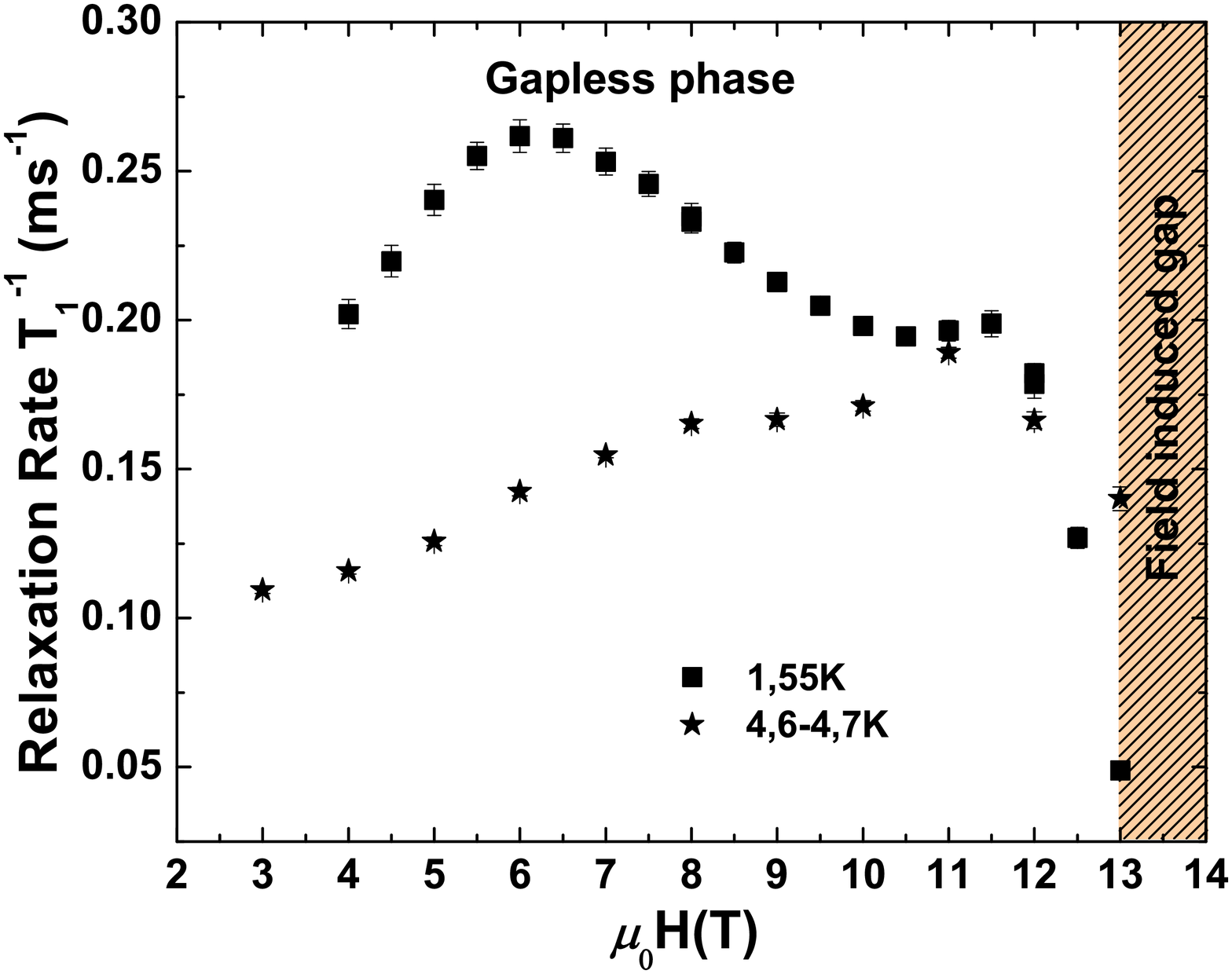}
\caption{ (Color online) Field dependence of the relaxation rate at $T=1.55\,$K and $T=4.6\,$K. The peak around $6\,$T at $T=1.55\,$K, may be a reminiscence at finite $T$ of the crossover between the VC and SDW$_2$ phases. $T_1$ drastically changes above $\mu_0H\approx 12\,$T, and above $\mu_0H=13\,$T the system is in the field induced gap phase. Between $12\,$T and $13\,$T it is plausible to search for the possible existence of a QN phase. }
\label{fielddep}
\end{figure}

\begin{figure}
\includegraphics[width=88mm]{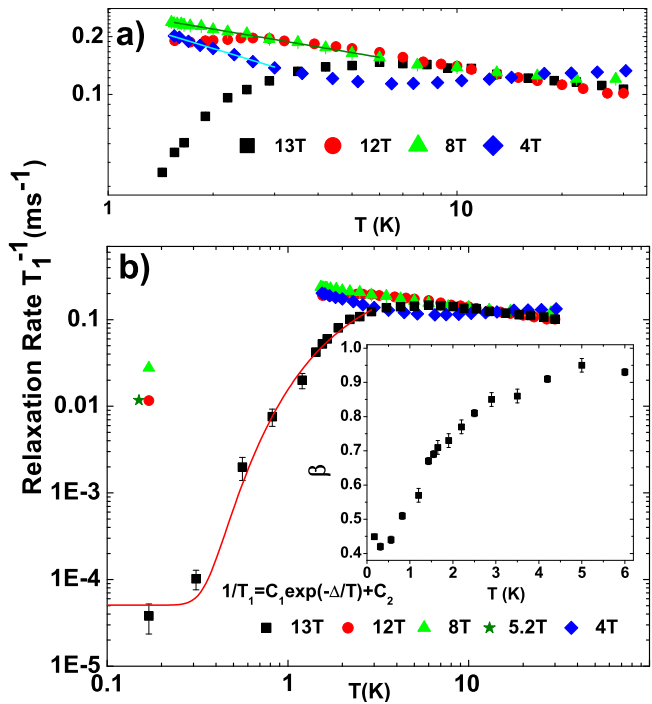}
\caption{ (Color online) (a) Temperature dependence of the spin lattice relaxation $T_{1}^{-1}$ in magnetic fields $\mu_0H=4, 8$ and $12\,$T.  In $4\,$T and $8\,$T relaxation rate obeys power law behaviour down to $1.55\,$K with critical exponents -0.55(2) and -0.27(1), respectively. At $12\,$T and $13\,$T broad maxima appear, but the drop in $13\,$T below $2.2\,$K is too steep to be assigned to a QN phase. (b) Additional low temperature (DR) measurements down to 125mK. 13T data below 2.9K were fitted to the function  $1/T_1=C_1\exp(-\Delta/T)+C_2$, where $\Delta=3.24(19)$K, with $C_1=0.396(47)$ms$^{-1}$ and $C_2=5.1(1.4)\times 10^{-5}$ms$^{-1}$.
Inset: $T$-dependence of the stretched exponent $\beta$, indicating a broad distribution of relaxation times at low temperatures.
}
\label{tempdep}
\end{figure}

For the measurements of the spin-lattice relaxation times $T_1$, a saturation-recovery pulse sequence was used. The first saturation pulse (4-5~$\mu$s)  rotates the nuclear magnetization in the $xy$ plane. After a waiting time $\tau_1$, the longitudinal magnetization recovered along $z$, the direction of the applied field, is  detected with a standard solid echo sequence with 3~$\mu$s $\pi/2$ pulses. Typical recovery curves are shown in Fig.~\ref{curves}. For applied field lower or equal to $12\,$T, at all temperatures, the recovery curve at the central peak position of the spectrum could be fitted to a single exponential giving a straightforward determination of $T_1$ (Fig. \ref{curves}a -- red circles). Above $12\,$T and below $5\,$K, the shape of the recovery changes  and can not be reproduced by a single exponential function (Fig. \ref{curves}a and c). This is certainly related to the occurrence of significantly different relaxation times for various spectral segments (Fig. \ref{curves}b), namely the left part of the spectrum (strongly coupled line) is relaxing slower than the right one. To take into account these different contributions of the two Li sites at the main peak position (see Fig.~\ref{spectra12T}), we used  a function of the general form:
\begin{equation}
\label{relaksacija}
1-\frac{M(\tau_1)}{M_{\infty}}=ae^{-(\tau_1/T_1)^{\beta}}+(1-a)f(\tau_1)
\end{equation}
where $M_{\infty}$ is the magnetization at equilibrium, $a$, the weight of the best coupled site, could be fixed to 0.72 in line with the analysis of the spectral shape in Fig.~\ref{spectra12T} and $f(t)$ is the residual relaxation associated to the least shifted site which could be determined independently by measurements on the right side of the spectrum (see Appendix B). The $T$-dependent stretched exponent $\beta < 1$ accounts for an increasing distribution of the $T_1$ values at low $T<5\,$K (see inset Fig.~\ref{tempdep}b). Such a distribution may be due to some local excitations which can be intrinsic property of chains or due to impurities (switching Cu with Li or Sb\cite{Dutton2012}) or defects (vacancies and chain edges). The upper limit for the impurity concentration is 1\% but they can significantly perturb a dozen of neighboring spins differently~\cite{Eggert2007, Eggert2011, Tedoldi1999},  leading to a distribution of relaxation times and the appearance of stretched exponentials~\cite{Shiroka2011}.

Field dependent $T_1$ measurements were performed at two temperatures, $1.55\,$K and $4.6\,$K  (Fig.~\ref{fielddep}) in the short range correlated phase as determined from previous heat capacity data~\cite{Dutton2012} and above magnetic transitions ($T_N \sim 0.7\,$K). At $1.55\,$K there is a peak in the relaxation rate ($1/T_1$) around $6\,$T, where the macroscopic magnetization reaches 20\% of its saturation value which disappears upon heating the system at $4.6\,$K closer to the paramagnetic state. This might be the finite temperature signature of the boundary between VC and SDW$_2$ phases which was theoretically predicted at 20\% of the saturated magnetization $M_s$~\cite{Sudan2009,HeidrichMeisner2009}.
At still higher fields, a quadrupolar nematic phase is expected above 70\% of $M_s$. The magnetization curve in that field region is very steep, and the field interval in which one may expect a QN phase is very narrow. The exact field value of the phase boundary between SDW$_2$ and QN  is still unspecified, however a qualitative change of the temperature dependence of $1/T_1$ should definitely mark this boundary. More specifically, from theory in strictly 1D systems with a magnetic transition only allowed at $T=0\,$K,  $1/T_1$ at very low temperatures  should follow a power law behavior $1/T_1 \propto T^{2\kappa-1}$ in both SDW$_2$ and QN phases, with $\kappa \in [0,1]$ being a field dependent Tomonaga-Luttinger parameter ~\cite{Sato2009,Sato2011}. Boundary between the two phases occurs at a field where $\kappa =1/2$, roughly when 70\% of saturation magnetization is reached. In SDW$_2$ phase ($\kappa < 1/2$) one therefore expects a power-law divergence of $1/T_1$ at low temperatures, while for the QN phase ($\kappa > 1/2$) a power-law drop towards zero is predicted. Above the saturation field, the system is gapped and $1/T_1$ should decay exponentially towards zero. With this in mind, we chose fields of $\mu_0H_0=4, 8, 12$ and $13\, \rm T$ to follow the temperature dependence of the relaxation rate (Fig.~\ref{tempdep}).

In $\mu_0H_0=4\,$T, where the VC phase is expected, $1/T_1$ shows traces of divergence down to $1.55\, \rm K$ with a critical exponent $1/T_1\propto T^{-0.55(2)}$ ($\kappa=0.24$).
In $\mu_0H_0=8\,\rm T$, where the SDW$_2$ phase is expected, $1/T_1$ shows a similar behavior down to $1.55\,$K with a critical exponent $1/T_1\propto T^{-0.30(1)}$ ($\kappa=0.38$).
Macroscopic magnetization measurements indicate that $\mu_0H_0=12\,\rm T$ is close to SDW$_2$/QN phase boundary. The temperature dependence of $1/T_1$ shows a broad maximum around $2.5\,$K which might indicate the proximity of the QN phase.
In $\mu_0H_0=13\,$T a maximum in relaxation rate occurs at much higher temperature, around 6K, but below 2.2K, the drop is too steep to be credited to a QN phase. If  we take into account DR measurements, the critical exponent is $1/T_1\propto T^{3.6}$, which would give a nonphysical value of $\kappa =2.3$. A more likely explanation, consistent with the formerly discussed static measurements, is that the magnetization is already saturated at $\mu_0H_0=13\,\rm T$, and the $1/T_1$ shows a gapped (exponential) behaviour. The $13\,$T data below 2.9K were then fitted to the function  $1/T_1=C_1exp(-\Delta/T)+C_2$, where $\Delta=3.24(19)$K,  $C_1=0.396(47)$ms$^{-1}$ and $C_2=5.1(1.4)\times 10^{-5}$ms$^{-1}$. The latter $T$-independent relaxation likely accounts for the contribution of a tiny amount of impurities or defects.
Hence, the temperature dependence of the spin-lattice relaxation time changes qualitatively and quantitatively in the narrow field range $12-13\,\rm T$, from a power-law divergence characteristic of the SDW$_2$ phase to a gapped behavior characteristic for saturation range.

Finally, the relaxation rate measured at a single temperature point $T=0.125\,\rm K$ in the field $\mu_0H_0=5.2\,\rm T$ gave a significantly longer $T_1$ ($86\, \rm ms$) than at $1.55\,$K, showing that there exists a maximum in the relaxation rate somewhere in between these two temperatures. As already noted, a peak in the heat capacity is observed at $700\,\rm mK$,\cite{Dutton2012} which, together with an anomaly in the susceptibility measurements,  may mark a phase transition causing the non-monotonous behavior of the relaxation rates.

\section{DISCUSSION}

\begin{figure}
\includegraphics[width=80mm]{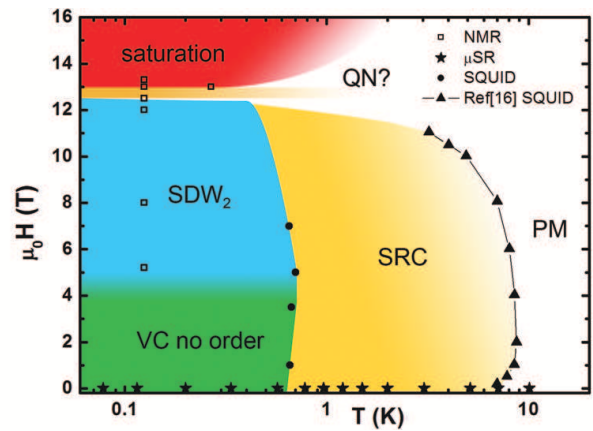}
\caption{ (Color online) Low-temperature phase diagram of LiCuSbO$_4$. Above 9K, the system is in the paramagnetic regime (PM - white). As the temperature is lowered short range correlations (SRC) among neighboring spins appear (yellow). However, from the width of NMR spectra in the temperatures down to 1.5K at various fields (4T, 8T, 12T, 13T), one can clearly dismiss any kind of 3D order (this point were omitted from the phase diagram for clarity). Below 0.7K, the system enters in a vector chiral (VC) regime with no static order in the fields up to $\approx 4$T. Above 4T a crossover to an ordered SDW$_2$ phase occurs, as evidenced by NMR (blue). In the high field and low temperature there is a narrow possibility for the existence of QN phase (orange), just below the saturation regime (red).}
\label{phasediag}
\end{figure}
In the following, we discuss the various low temperature phases of LiCuSbO$_4$ in the light of known data from the present study and Ref.~[\onlinecite{Dutton2012}] and summarize our current understanding in the tentative phase diagram  shown in Fig.~\ref{phasediag}. To label the various field induced phases, we rely on the theoretical work by Sudan \emph{et al}~\cite{Sudan2009}. From the latter study of frustrated spin chains, in LiCuSbO$_4$ with $J_1/J_2=-2.22$, the  magnetic excitations of $p=2$ bound spin flips should govern the phase diagram.  Several field-induced ground states are predicted: (i) a vector chiral phase (VC) for $M/M_{sat}<0.15$, (ii) an SDW$_2$ phase for $0.15<M/M_{sat}<0.7$, and (iii) a QN phase for $0.7<M/M_{sat}<1$. Using the NMR shift at 0.125~K from Fig.~\ref{saturacija} as a measure of the magnetization, one estimates that $M/M_{sat}=0.15$ occurs at $\mu_0H\approx 5\, \rm T$, $M/M_{sat}=0.7$ occurs at $\mu_0H\approx 12\, \rm T$, and $M=M_{sat}$ occurs at $\mu_0H\approx 13\, \rm T$.

{\it Vector chiral phase.} The main result obtained from $\mu$SR measurements is that in LiCuSbO$_4$ no 3D magnetic ordering occurs down to $T=30\, \rm mK$ in zero field, which is a comparative advantage to all known frustrated 1D chains. The possible reason lies in the crystal structure in which hard magnetization axes of adjacent chains are not parallel~\cite{Dutton2012}, thus preventing 3D ordering down to the lowest temperatures. This low-field low-temperature phase (labeled green in Fig.~\ref{phasediag}) is consistent with the theoretical prediction of vector chiral (helical) magnetic ground-state\cite{Sudan2009}. Here no static ordering can be stabilized by magnetic field  and the dominant VC correlation functions are faster than characteristic $\mu$SR (and NMR) timescale. Looking for the boundary between the VC phase and the predicted SDW$_2$ we notice that the susceptibility curves in 1 and 3.5~T (Fig.~\ref{Chi_lowT}) show the same shape with a minimum close to 0.7~K while at higher fields the shape evolves suggesting a distinct magnetic phase. Additionally, the heat capacity curves shown in in Figure 3b of Ref. [\onlinecite{Dutton2012}] are similar at 0 and 2~T and evolve gradually only above 4~T with the growing of a peak around $T=0.7$~K. Therefore the boundary between the two phases seems to occurs between 4 and 5 T. The gradual evolution of the heat capacity suggest rather a crossover than a sharp transition from the VC phase to the SDW$_2$ one.

{\it SDW$_2$ phase.} The NMR spectra measured at $T=125\, \rm mK$ in $\mu_0 H=5.2,$ and $8\, \rm T$ (Fig.~\ref{DRspectra}) clearly show that magnetic ordering has occurred. This is confirmed by specific heat measurements where a peak at $T=0.7\, \rm K$ clearly indicates a magnetic transition in this intermediate field range.  NMR relaxation measurements at $\mu_0 H=8\, \rm T$ above the transition temperature show the temperature dependence $1/T_1\propto T^{-0.30(1)}$ consistent with a SDW$_2$ phase (negative exponent less than 1).
These results give strong indication that in the medium field range between $\mu_0 H \approx 5,$ and $12\, \rm T$ the theoretically predicted SDW$_2$ phase is indeed stabilized by the magnetic field.

{\it Possible quadrupolar nematic phase.}  NMR relaxation measurements at $\mu_0 H=12\, \rm T$ show a negative slope above $T=2.5\,\rm K$, consistent with SDW$_2$ phase, but below $T=2.5\,\rm K$ the slope changes to positive which could be consistent with a QN phase (positive exponent less than 1). This behavior may point to the proximity of the crossover between SDW$_2$ and QN phases. Contrary to the SDW$_2$ phase, the QN phase should not broaden the NMR spectrum, which was recently pointed out by Orlova et. al. \cite{Orlova2017} in LiCuVO$_4$ just below the saturation field. Inspection of Fig.~\ref{DRspectra} shows that the spectrum at $\mu_0H = 12\, \rm T$ is narrower than the spectra at $\mu_0H=5.2\, \rm T$ and $8\, \rm T$, but still broader than the spectra at higher fields. This again points to the proximity of QN phase, yet still not reached. The spectrum at
$\mu_0H = 12.5\, \rm T$ is still narrower, maybe consistent with the QN phase. From the line shift the magnetization $M/M_{sat}=0.85$, is theoretically in the QN phase  ($M/M_{sat}>0.7$) but the QN phase could be pushed to even higher magnetization for systems with hard-axis anisotropy~\cite{HeidrichMeisner2009}. NMR relaxation measurements down to the lowest temperatures may help deciding whether the QN phase is realized at this field. In any case the lower boundary of the QN phase should be close to 12.5~T. On the other hand at $\mu_0H = 13\, \rm T$ the full saturation is reached, giving the upper boundary for the possible QN phase.

{\it Saturation regime.}
When the saturation is reached one expects a gap in the excitation spectrum. This is confirmed by our relaxation measurements in the dilution refrigerator shown in Fig.~\ref{tempdep}.b which yield the gap value $\Delta\approx 3.2\,\rm K$ at 13~T. A theoretical model for the isotropic case \cite{Kecke2007} predicts the opening of the gap (ie the entrance in the saturation regime) when the ratio of the spin magnetic energy $\epsilon_s=g\mu_B\mu_0H_s$  and the NNN interaction $J_2$ reaches $\epsilon_s/J_2\approx0.5$ which is in excellent agreement with our case (for g=2.21\cite{Dutton2012} in field of $\mu_0H_0=13$T $\epsilon_s=18.2$K while $J_2=34$K).

Grafe et al.\cite{Grafe2017_2} recently reported an extensive experimental and theoretical research on LiCuSbO$_4$. NMR measurements were performed up to 16T on an unoriented powder sample, in a temperature range down to 2K, preventing them from investigating the $T\rightarrow$0K  phase diagram. Nonetheless in the temperature and field range common to our study, the NMR results are consistent. In particular at 13~T they also observed a gapped behavior of $T_1$, with a gap value $\Delta\approx 2.4$K which is similar to our value. The main discrepancy with the present study is the determination of the saturation field which is of great importance in order to put an upper boundary on QN phase. In ref.~\onlinecite{Grafe2017_2}, the magnetization curve obtained from bulk measurement using pulsed fields at 0.45~K did not reach saturation up to the highest measured field of 20~T, which was interpreted as a signature of in plane exchange anisotropy ($J_1^x \neq J_1^y$). On the contrary our NMR shift data point to a well defined saturation field close to $13\, \rm T$ and do not demand modifications of the magnetic model.

For the other QN candidate, LiCuVO$_4$, where high quality single crystals are available, the NMR spectra could be followed in detail and match the two criteria for a QN phase, shift variation and minimal width~\cite{Orlova2017}. The quality of the sample needed and the detailed information on the hyperfine tensors are available for this compound. Yet the absence of $T_1$ measurements due to the high saturation field remains a severe bottleneck to the definitive proof of a QN phase.


So far we interpreted our results guided by the theory for isotropic $J_1-J_2$ chains while our compound possesses a small easy plane anisotropy ($\approx0.8$)~\cite{Dutton2012}. A detailed analysis of phase diagram in such anisotropic case is given in Ref. [\onlinecite{HeidrichMeisner2009}]. Easy plane anisotropy favors vector chiral phase, but if the anisotropy is not considerably pronounced, the QN phase still may exist in the high field regime.
This is possible reason why the QN phase is restricted to so narrow field range just below the saturation.

\section{SUMMARY}

LiCuSbO$_4$ features a very rich phase diagram with the possibility of having an exotic, not yet observed QN phase. In the limit $T\rightarrow 0$ the ground state depends heavily on the applied magnetic field. In zero field no 3D ordering is observed down to $30\,$mK, although $\mu$SR shows traces of short range correlations. This low field phase is identified to a theoretically predicted vector chiral phase.  The crossover between the vector chiral and the higher field spin density wave SDW$_2$ phase occurs between $4\,$T  and $5\,$T. Indeed, above this crossover, the broadening of the NMR spectra reveals an induced 3D magnetic ordering on the scale of $0.75\mu_B$. Further, the field dependent NMR shift data at $125\,$mK enable a precise determination of the saturation field, which turns out to occur sharply and close to $13\,$T.
 
The NMR spectrum is still broad at $12\,$T, contrary to the predictions for the QN phase. However, the spectrum at $12.5\,$T is narrower and featureless, while its shift indicates that the magnetization is not yet saturated. This is one signature of a QN phase, namely  the absence of rotation symmetry breaking just below the saturation field. Hence, combined with the change of behavior of the $T$-dependence of the relaxation rate, we point to the possible existence of a  QN phase in a narrow field range between $12.5\,$T and $13\,$T. 

Our study provides a strong motivation to refine the investigations in a still very accessible field range as compared to other $J_1-J_2$ candidates and look for the final signatures of the long sought QN phase in spin chain systems. If the direct observation of the elusive order parameter of a QN phase is cumbersome, two theoretical predictions on experimental observables indeed pave the way: NMR relaxation measurements should be performed  down to very low temperature following Ref.\onlinecite{Sato2009,Sato2011}, as pointed out in Ref.~\onlinecite{Mila2017} and  ESR might reveal excitation of bound magnon pair typical of a QN phase \cite{Furuya2017}.

\section*{ACKNOWLEDGMENTS}
 This work was
supported by the Croatian Science Fundation (HRZZ) Grant no. IP-11-2013-2729, the French Agence Nationale de la
Recherche under “SPINLIQ” Grant No. ANR-12-BS04-0021, the  Universit\'{e} Paris-Sud under "PMP" MRM Grant, and the
 R\'{e}gion Ile-de-France under "MAGMAT" SESAME Grant.
The authors thank Ivica \v{Z}ivkovi\'c for complementary measurements.

\appendix

\section{Angle dependent oriented powder spectra}
\begin{figure}[b!]
\includegraphics[width=88mm]{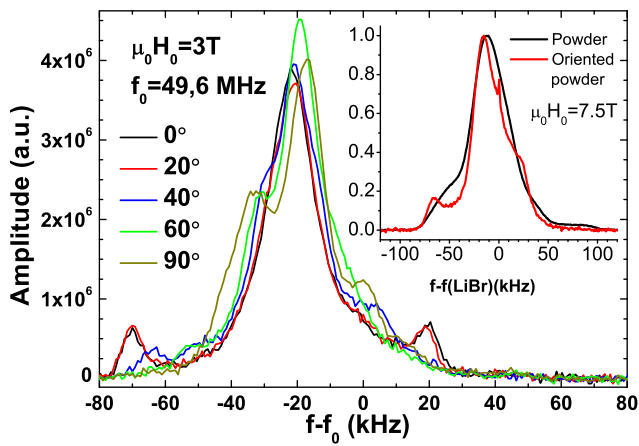}
\caption{(Color online) Room temperature spectra at $\mu_0H=3\,$T for different orientations of the applied field in the $bc$ plane. The quadrupolar splitting is the broadest when the applied field is parallel to the hard axis ($0 ^{\circ}$). Inset: Unoriented vs. oriented powder spectrum measured in $\mu_0 H=7.5\, \rm T$. The small peak at zero shift comes from the LiBr reference signal.}
\label{rotation}
\end{figure}

After the hardening of the mixture of epoxy glue and the powder sample, the orientation was checked by measuring the angular dependence of the NMR spectra in the magnetic field of 3T as shown in Fig.~\ref{rotation}. The quadrupole satellites of the strongly coupled line are most pronounced when the magnetic field is applied in the direction of the hard axis, which means that the eigenvector corresponding to the electric field gradient tensor component $V_{zz}$ points in the direction of the hard axis (within an experimental error of $\approx 10^\circ$). As soon as the applied field is not parallel to the hard axis, the powder spectra narrow down and the spectral analysis becomes cumbersome.
The inset of Fig.~\ref{rotation} shows the unoriented vs. oriented powder spectrum in the magnetic field of 7.5T. In the unoriented powder spectrum (black line) the quadrupole satellites cannot be resolved. For consistency we made a couple of oriented powder samples which have shown identical spectra when the applied field was parallel to the hard axis.

\section{Relaxation fit parameters}

As mentioned in section VI, at the peak position the spectra have contributions from both the strongly and weakly coupled lines, with different relaxation times. To determine the contribution of the weakly coupled line which corresponds to the function $f(\tau_1)$ in Eq. (\ref{relaksacija}) we measured the relaxation on the right side of the spectra at 30\% of the peak value where the contribution of the strongly coupled site is negligible. The relaxation curves of the weakly coupled line reveal two different relaxation regimes: a fast one with a relaxation time $T_{1b}$ shorter than $T_1$ of the strongly coupled site, but of the same order of magnitude, and a long one $T_{1c}$ which is several orders of magnitude longer than $T_1$ and $T_{1b}$ at low temperatures. Our main goal was to fix $T_{1b}$ so we could isolate $T_1$ in the Eq. (\ref{relaksacija}). We used the function $f(\tau_1)$ (for the weakly coupled site):
\begin{equation}
\label{relaksacija2}
f(\tau_1) =b e^{-(\tau_1/T_{1b})^{\beta_b}}+ c e^{-(\tau_1/T_{1c})^{\beta_c}}.
\end{equation}
Having set $(1-a)=0.28$ in Eq.~\ref{relaksacija}, the weights $b=\frac{0.22}{0.28}$ and $c=\frac{0.06}{0.28}$ were determined to fit the contributions of the fast and slow relaxation processes. The stretch parameters $\beta_b$ and $\beta_c$ were fixed to 1 above 1.55K and left as free parameters at lower temperature measurements.

Spectra taken in $\mu_0H_0=$12T at different temperatures show indeed that the spectral shape does not change appreciably between 1.55 K and 30 K, and we take the coefficients $a$, $b$ and $c$ as temperature independent. The spectral shape does not change either between 12T and 13T at 1.55K, and we conclude that these parameters are field independent as well (for high fields). The parameter $M_{\infty}$ was taken from the last measured point at $\tau_1>7\times T_1(T_{1b})$, limited by the repetition time which had to be reasonably short in order to make measurements possible. The total signal did not relax completely between two acquisitions, but this affects only the value of the longest relaxation time $T_{1c}$. Since $T_{1c}$ is an order of magnitude longer than $T_1$ and $T_{1b}$, it turns out that its exact values are not critical for the determination of $T_1$.  The values of $T_{1b}$ was determined for each temperature and fixed before the $T_1$ in Eq. (\ref{relaksacija}) was calculated. The stretch parameter $\beta_b$ ranges from 0.51 at lowest temperature to 1 at 1.55 K.

\bibliography{Spin_Sytems}
\end{document}